\documentclass[12pt]{iopart}
\usepackage{iopams}  
\usepackage{amstext}
\usepackage{graphicx}
\usepackage{subcaption}
\usepackage{hyperref}
\usepackage{latexml}
\iflatexml
\else
\usepackage{eso-pic}
\fi

\begin{document}
\def\copyrightdoi{10.1088/1402-4896/ad4ca4}
\def\copyrightref{Physica Scripta}
\def\copyrightyear{2024}
\def\copyrighttext{
This is the Accepted Manuscript version of an article accepted for publication in \copyrightref. IOP Publishing Ltd is not responsible for any errors or omissions in this version of the manuscript or any version derived from it.  The Version of Record is available online at \href{http://dx.doi.org/\copyrightdoi}{\copyrightdoi}.

This Accepted Manuscript is available for reuse under a \href{https://creativecommons.org/licenses/by-nc-nd/3.0/}{CC BY-NC-ND} licence after the 12 month embargo period provided that all the terms of the licence are adhered to.

Although reasonable endeavours have been taken to obtain all necessary permissions from third parties to include their copyrighted content within this article, their full citation and copyright line may not be present in this Accepted Manuscript version. Before using any content from this article, please refer to the Version of Record on IOPscience for full citation and copyright details, as permissions may be required. All third party content is fully copyright protected, unless specifically stated otherwise in the figure caption in the Version of Record.
}
\iflatexml
\textit{\copyrighttext}

\else
\AddToShipoutPictureBG*{%
	\AtPageUpperLeft{
		\hspace*{\dimexpr0.5\paperwidth\relax}
		\makebox(0,-5cm)[c]{\framebox{\parbox{0.8\paperwidth}{%
			\footnotesize \copyrighttext}
		}}%
	}%
}%
\fi
\title[Superconducting Electronic Linearity Calculation Method]{Superconducting Electronic Device Response Linearity Calculation Method Using Discrete Fourier Analysis}

\author{Nikolay V. Kolotinskiy, Victor K. Kornev}

\address{Lomonosov Moscow State University, Leninskie Gory, 1, b.2., Moscow, Russia}
\ead{kolotinskij@physics.msu.ru}

\begin{abstract}
A highly effective method of the device linearity calculation on the stage of the device development is worked
out and reported. The method allows expressing the linearity in terms of the achievable spurious-free
dynamic range (SFDR) of the created devices, in particular superconductive electronic devices, and therefore
can be easy correlated with the obtained experimental data. The algorithm scheme and the method accuracy
are considered and discussed in detail.
\end{abstract}

\vspace{-2pc}
\submitto{\PS}
\vspace{1pc}

\section{Introduction}

Spurious-free dynamic range (SFDR) \cite{IEEE_LIN} is key characteristic of all active devices, such as amplifiers, active antennas and so on, which are used in communication systems. SFDR is a measure of signal purity and can be defined as the ratio of the maximum amplitude of the signal component at the device output to the next larges value of noise or amplitude of intermodulation (nonlinear distortion) component. As far as rms value of noise can be reduced with decrease in frequency band, fundamental restriction on the attainable maximum value of SFDR is imposed by the device linearity/nonlinearity property. Experimental evaluation of SFDR of a fabricated device can be performed using the standard single-tone or two-tone measurement techniques \cite{IEEE_LIN}. However, usually one needs to have an estimate of the device linearity at design stage. Depending on specific task, the designer may use Pearson’s chi-squared test for linear fitting \cite{Prokopenko2013} or the first-derivative method \cite{Ferrante2023} to evaluate deviation of the output signal under study from an optimal linear characteristic. These methods are useful for solving optimization problems, but their use cannot be easily correlated with the subsequent experimental values of the device characteristics.
 
This article describes an effective method for numerical evaluation of the output signal linearity of the devices under development in terms of SFDR. This method was successfully used in solving optimization problems and in development of a number of active superconductor devices including differential Josephson-junction circuits, bi-SQUIDs, active electrically small antennas (ESA), and other devices \cite{Kolotinskiy2024,Kornev2017_ESA,Kornev2020_SUST}.

\section{Spurious-free dynamic range as a measure of linearity}
In the absence of noise, as it usually occurs in numerical simulations, SFDR is completely determined by the only device linearity characteristic and hence can be used as the measure of device linearity. Therefore, the linearity can be calculated in the way used to measure SFDR in accordance with either one- or two-tone measurement technique, i.e. with applying an input signal in the form of a one-tone signal at frequency $\omega_1$ or two tones with equal amplitudes at close frequencies $\omega_1$ and $\omega_2$. Then, the output signal spectrum is measured to calculate the response linearity in decibels with using the following formula \cite{IEEE_LIN}:
\begin{equation}
	\label{lin_def}
	\text{Lin} = 20\cdot \log \left(a_1/\max{\left\{a_k\right\}}\right)
\end{equation}
where $a_1$ is the amplitude of fundamental ton(s) and $a_k$ is the amplitude of either the signal harmonics at frequencies $k\omega_1$, where $k=2,3,\dots$, (in the one-tone technique) or the intermodulation components at frequencies $n\omega_1\pm m\omega_2$ with $k=n+m=2,3,\dots$ (in the two-tone technique). The factor ``20'' must be changed to ``10'' when power spectrum is used since the measured power spectrum components are proportional to $a_1^2$ and $a_k^2$, respectively.

\begin{figure}[t]
	\centering
	\includegraphics[width=9cm]{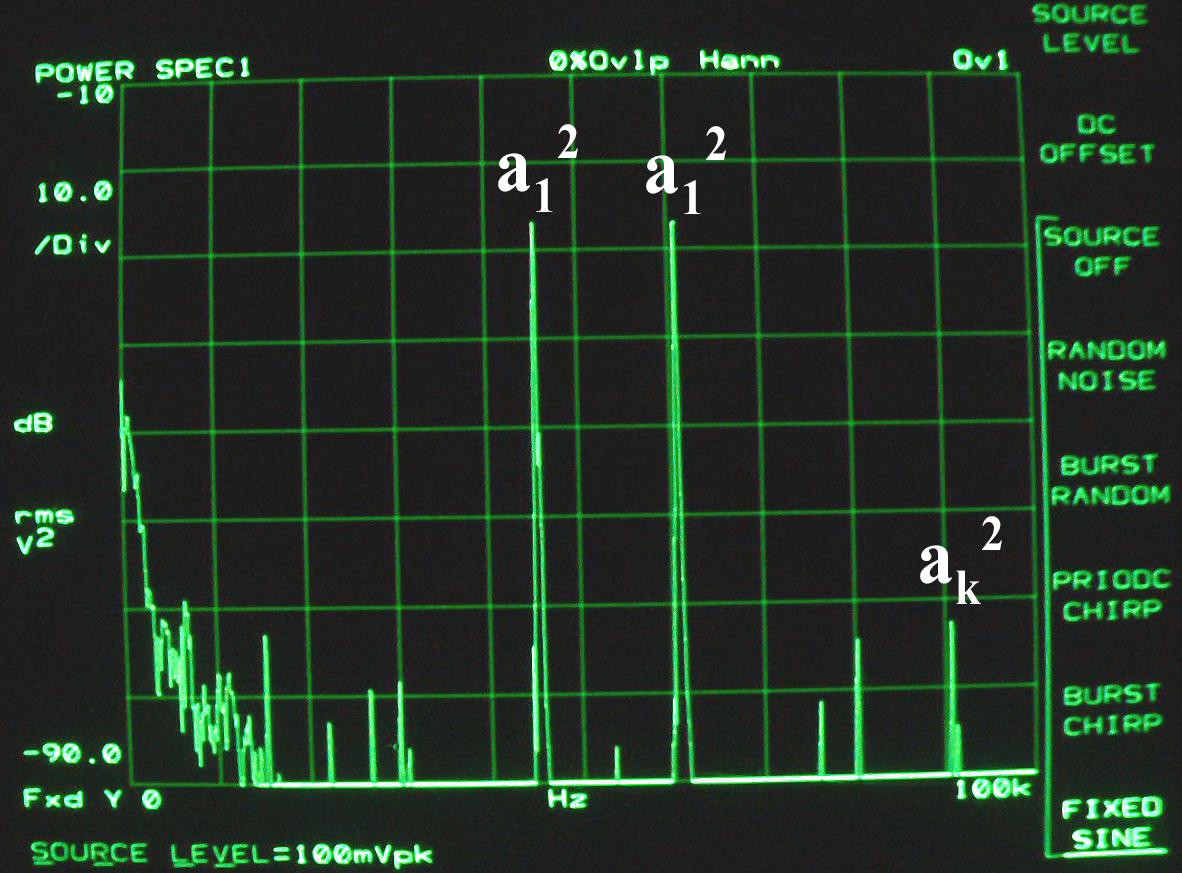}
	\caption{\label{fig:SFDRExample}The measured power spectrum of voltage response of a 128-cell serial bi-SQUID array on the applied two-tone ($45\text{ kHz}$ and $60\text{ kHz}$) input signal. Here $a_1^2$ is the fundamental tone power, $a_k^2$ is the power of the maximal intermodulation component of the output signal. Since noise level is much less than the observed intermodulation components, both the linearity and SFDR are the same and as high as about 45 dB. From \cite{Kornev2020_SUST} with permission.}
\end{figure}

Figure~\ref{fig:SFDRExample} shows an example of the measured power spectrum of the device output voltage response on the applied two-tone input signal. As far as noise level at frequencies $f > 15\text{ kHz}$ can be considered as negligibly small, SFDR is determined completely by the device linearity and therefore both the linearity and SFDR in this frequency range can be expressed in dB by ratio of the fundamental tone power $a_1^2$ to the largest intermodulation component power $a_k^2$. Since the shown power spectrum is measured directly in dB, the linearity Lin and SFDR are just the difference of these spectral components, both expressed in dB, that gives about $45\text{ dB}$ as seen from Fig.~\ref{fig:SFDRExample}.

In fact, it should be noted that SFDR is primarily a measure of signal purity, and not just linearity. This is because it takes into account both noise (unwanted spurious signals) and distortion caused by nonlinearity, and thus essentially represents the effective dynamic range of a system. However, unlike noise, which can have a lower RMS value with a decreasing frequency band, nonlinear distortion places fundamental limitations on the SFDR, making it a measure of linearity in itself.

\section{Algorithm}
The main idea of the method for determining the linearity from the calculated data is quite simple. The input data for the method is the output signal of the device $V_\text{signal}(F)$ (where $F$ is a parameter, for SQUIDs system it could be, \textit{e.g.},  an external magnetic flux) calculated by some program (PSCAN \cite{Polonsky1991,Shevchenko2016}, WRSpice \cite{WRSpice}, etc.), and the result of applying this method should be the linearity of this signal in terms of SFDR.

Let us consider this method in details (see the flowchart in figure~\ref{fig:algorithm}):
\begin{figure}[b]
	\centering \includegraphics[width=15cm]{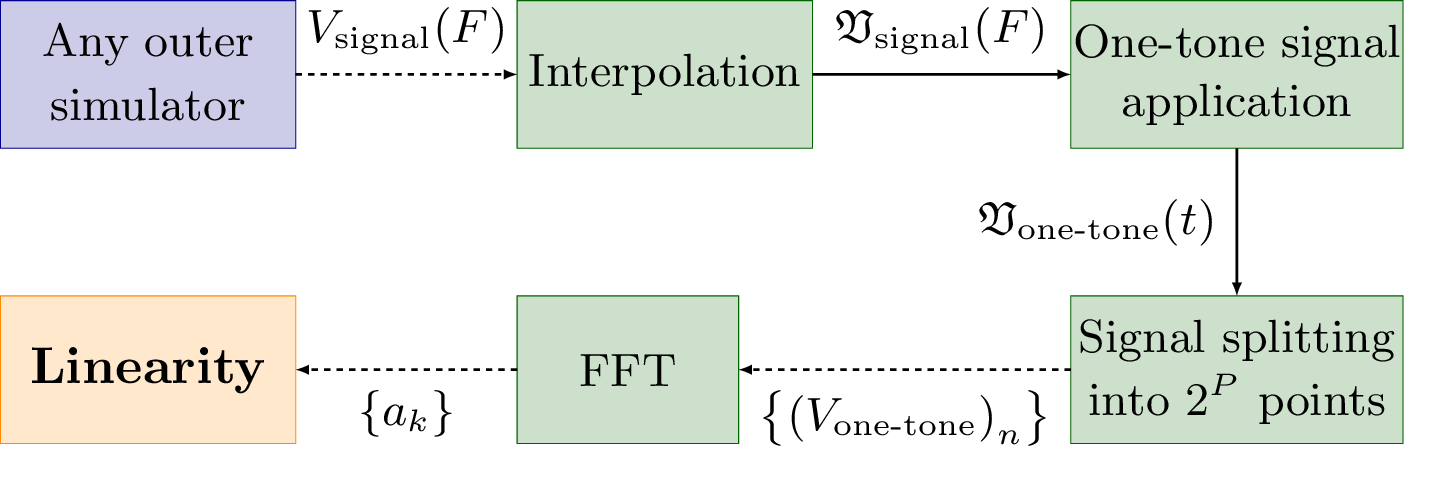}
	\caption{\label{fig:algorithm}Flowchart for linearity calculation algorithm. $V_\text{signal}(F)$ -- original discrete signal (calculated by outer program), for example, voltage-flux characteristic; $\mathfrak{V}_\text{signal}(F)$ -- continuous signal; $\mathfrak{V}_{one-tone} (t)$ -- output signal of device, when a one-tone signal applied as input (time-domain signal); $\left\{\left(V_\text{one-tone}\right)_n\right\}$ -- FFT-ready array; $\left\{a_k\right\}$ - set of amplitudes of harmonics. Dashed lines represents discrete functions (or arrays), solid -- continuous functions.}
\end{figure}
\begin{enumerate}
\item Suppose we have a calculated parameter-domain signal $V_\text{signal}(F)$, which is a set of $N$ points of the form $\left(V_\text{signal}\right)_n (F_n)$; in cases of SQUIDs systems this parameter-domain signal is, for example, voltage-flux characteristic. In general, the external parameter  $F_n$ may not be distributed evenly. However, for simplicity, we will assume that they were calculated with a constant step $\Delta F$. Note that this signal does not yet contain the time-dependence. This is the device's response function and not the quasi-harmonic signal under investigation. The modification of the method in the event that the simulation results in time dependence $V_\text{signal}(t)$ will be considered in the \ref{sec:ModMethod}.
\item Before applying a sinusoidal signal to our response function, we must move from a discrete consideration to a continuous one. Since we postulate that the response is accurately measured  at points $F_n$ (with an allowance for the relative error $\varepsilon$), it is logical to choose an interpolation method to describe the system between points $F_n$ and $F_{n+1}$. Moreover, the use of smoothing methods (which do not guarantee that the resulting function will be equal to the original signal at the measured points) would lead to an obviously incorrect result. Indeed, if we smooth our response with a polynomial $P(x)$ of degree $m$, then further analysis would show the linearity of this polynomial rather than the original response. The only way to determine the linearity of the original response would be through the accuracy of the smoothing, which is precisely what we wished to avoid. Therefore, for reconstructing the response, we will employ an interpolation technique using polynomials of degree $m$. This degree $m$ should be sufficient for the intended description of our system. For most SQUID systems, it is sufficient to choose $m = 3\text{ to }6$. Denote this reconstructed response as $\mathfrak{V}_\text{signal}(F)$, which still represents a signal in the parameter domain.
\item Using $\mathfrak{V}_\text{signal}(F)$, we can obtain the device's output time-domain signal when a one-tone signal with a certain frequency $\omega$ is applied to its input: $\mathfrak{V}_{one-tone} (t) = \mathfrak{V}\left(A\sin\left(\omega t\right) + A_0\right)$, where $A$ and $A_0$ are some constants (amplitude and offset, respectively). The choice of these constants is determined entirely by the choice of the part of the response we wish to investigate. The choice of frequency (in relative units) depends on the number of points $2^P$ used for further Fourier analysis. As we have the flexibility to change each of these quantities, we can set the frequency arbitrarily (if we assume that the response does not depend on frequency; otherwise, we need to use the modified method described in the \ref{sec:ModMethod}). Note that $\mathfrak{V}_{one-tone}(t)$, defined in this way, is definitely continuous in $t$.
\item From the dependence $\mathfrak{V}_{one-tone}(t)$, we can derive $2^P$ points $\left\{\left(V_\text{one-tone}\right)_n\right\}$ in the interval $t\in\left[0;T\right]$. $T$ should be large enough to include many periods, i.e. $T \gg 1/\omega$. The resulting array $\left\{\left(V_\text{one-tone}\right)_n\right\}$ can then be processed by Fast Fourier Transform (FFT), which will yield a set of amplitudes $\left\{a_k\right\}$, where $a_1$ represents the amplitude of the fundamental tone and $a_k$ ($k \ge 2$) is the amplitude of harmonic components of the signal.
\item Finally, using the array $\left\{a_k\right\}$, we can determine the linearity by the formula \eref{lin_def}.
\end{enumerate}
In \ref{sec:Example} example of this method implementation for analysis are shown.

\section{\label{sec:discus}Discussion of the method}

\subsection{Requirements for the input data of the method (signal)}
We assume that the input data ($V_\text{signal}$) were obtained from some program for modeling superconducting electronics devices. To make the method work, it is necessary to ensure that the simulated signal adequately describes the response function of the device. The term ``adequately'' can be understood as follows \cite{Kokoshkin2016}: the number of points in the simulated signal allows us to make an unambiguous assumption about the amplitude and shape of that signal.

\subsection{FFT accuracy}
FFT methods are characterized by high accuracy. Moreover, when using accurate FFT methods (such as, \textit{e.g.}, the Cooley-Tukey FFT algorithm \cite{Cooley1965}), the introduced error depends only on the floating-point accuracy $\varepsilon_\text{float}$ and is approximately equal to $\varepsilon_\text{float}\sqrt{P}$, where $P$ is the logarithm with base two of the number of FFT points used \cite{Schatzman1996}. This error can be estimated to be $10^{-10}$. Thus, the introduced error is considered to be insignificant.

\subsection{The issue of selecting FFT parameters}
We need to choose three parameters when processing an FFT signal: the frequency $\omega$, the number of points $2^P$, and the time $T$. By converting from a discrete function to a continuous one, we can avoid the standard problems associated with FFT analysis and instead use a rectangular window. For proper consideration, we just need to formulate the following requirements for $T$, $2^p$, and $\omega$: an arbitrary integer number of cycles should fit within $T$, so $T$ should be equal to $M \times 2\pi / \omega$, where $M$ is an integer; and for harmonic detection purposes, it's reasonable to require that $2^P \gg \left(5\text{ to }10\right)\times\omega$.

\subsection{The issue of the effect of input signal inaccuracy on the obtained linearity value}
Taking into account the transfer of inaccuracy (error) of the input signal (the so-called fuzzy data) into the error of the interpolation polynomial is a non-trivial problem \cite{Valenzuela2011}. Its solution may be using fuzzy splines \cite{Kaleva1994}. Denote the relative error in the original data by $\varepsilon$ and the resulting error by $\varepsilon_\text{int}$.

Another factor affecting the inaccuracy in the obtained linearity value is the transformation of the error $\varepsilon_\text{int}$ during the FFT. This error $\varepsilon_\text{lin}$ is the result error of our method. Although FFT can be considered as linear, this issue is often resolved by using experimental mathematics (through numerical experiments), especially for estimating the upper bound \cite{Calvetti1991}.

The key idea behind the numerical experiment to estimate $\varepsilon_\text{int}$ is as follows: we conduct a large number of measurements of the test signal (a linear signal with random errors with a relative magnitude of $\varepsilon$), obtaining a certain distribution of linearity values. We then consider the relative error $\varepsilon_\text{lin}$ as $\sqrt{\sigma^2} / M$, where $\sigma^2$ is the dispersion and $M$ is the mean value of this distribution.

\begin{figure}
	\centering\includegraphics[width=9cm]{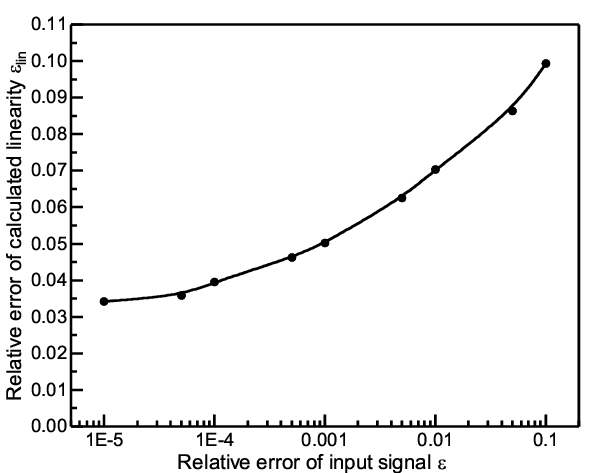}
	\caption{\label{fig:Error} Relative linearity error $\varepsilon_\text{lin}$ versus relative error $\varepsilon$ analyzed signal data modeled with using numerical simulation software. The data obtained through numerical experiment  (the number of measurements is $10^3$, the polynomial degree $m=6$ and the number of FFT points is $2^{12}$).}
\end{figure}

Figure~\ref{fig:Error} shows results of such a numerical experiment (the number of measurements is $10^3$, the polynomial degree $m=6$ and the number of FFT points is $2^{12}$). In particular, at a relative error of $\varepsilon=10^{-3}$, the relative error in determining the signal linearity is $\varepsilon_\text{lin}=0.05$, which is equal to $4\text{ dB}$ at the signal linearity of about $80\text{ dB}$.

It should be noted that the inaccuracy of the original signal (due to both the numerical noise $\varepsilon$ and the specially introduced sources to model thermal noise) affects the linearity and decreases its value. This is not a method error but rather a reflection by the method of the actual decrease in the values of linearity: the presence of noise affects the amplitude of harmonics and consequently lowers linearity in terms of SFDR.

\section{Conclusion}
In such a way, a highly effective method of the device linearity calculation based on the using numerical simulation data obtained on the stage of the device development is worked out and reported here. In the algorithm, linearity error is as low as about $4\text{ dB}$. The method allows expressing the linearity in terms of the attainable SFDR of the device under development and therefore can be easy correlated with the obtained experimental data. 

\ack
This research was supported by Russian Science Foundation (RSCF) through grant no. 19-72-10016-P.

N.V.K. is grateful to T. Filippov for his advice to consider this issue in more detail, which led to the writing of this article.

\section*{CRediT author statement}
Conceptualization: N.V.K. and V.K.K.; Formal analysis: N.V.K.; Funding acquisition: N.V.K.; Investigation: N.V.K.; Methodology: N.V.K. and V.K.K.; Project administration: N.V.K.; Software: N.V.K.; Supervision: N.V.K. and V.K.K.; Validation: N.V.K. and V.K.K.; Writing -- original draft: N.V.K.; Writing -- review \& editing: N.V.K. and V.K.K.

\section*{Conflict of interest statement}
The authors declare no conflict of interest. The funders had no role in any part of the manuscript preparation process.


\section*{References}
\bibliographystyle{iopart-num}
\bibliography{Kolotinskiy_LinearityMethod.bib}

\appendix
\section{\label{sec:ModMethod} Modification of the algorithm if a one-tone signal is present in the initial signal.}
Let us consider a modified algorithm in which the initially signal already contains an applied one-tone signal.
\begin{enumerate}
	\item Suppose we have a calculated signal $V_\text{one-tone}(t)$, which is a set of $N$ points of the form $\left(V_\text{one-tone}\right)_n (t_n)$. In general, the $t_N$ may not be distributed evenly, however, for simplicity, we will assume that they were calculated with a constant step $\Delta t$. 
	\item Although we already have a time-dependent signal, we still need to convert our signal into a continuous one. Similarly to the original algorithm, we reconstruct the signal between points $t_n$ and $t_{n+1}$ using interpolation methods, denote the reconstructed signal as $\mathfrak{V}_\text{one-tone} (t)$.
	\item Continuation of algorithm is the same: 
	\begin{enumerate}
	\item From the dependence $\mathfrak{V}_{one-tone}(t)$, we can derive $2^P$ points $\left\{\left(V_\text{one-tone}\right)_n\right\}$ in the interval $t\in\left[0;T\right]$, where $T$ must be large enough to include many periods, i.e. $T \gg 1/\omega$. The resulting array $\left\{\left(V_\text{one-tone}\right)_n\right\}$ can then be processed by Fast Fourier Transform (FFT), which will produce a set of amplitudes $\left\{a_k\right\}$, where $a_1$ represents the amplitude of the fundamental tone and $a_k$ ($k \ge 2$) is the amplitude of harmonics of the signal.
	\item Finally, using the array $\left\{a_k\right\}$, we can determine the linearity using the formula \eref{lin_def}.
\end{enumerate}
\end{enumerate}

Since the methods used are the same (only their order has changed), all the reasoning about the accuracy of the method presented in \sref{sec:discus} remains unchanged as well, except for the considerations about selecting FFT parameters. A more complex consideration of using windows is beyond the scope of this appendix. However, taking into account the known frequency of the one-tone signal, we can always choose a time range $T$ such that an integer number of periods fit into it.

\section{\label{sec:Example} Example of using the method}
\begin{figure}

\subcaptionbox{\label{fig:biSQUID}}{\centering\includegraphics[width=6cm]{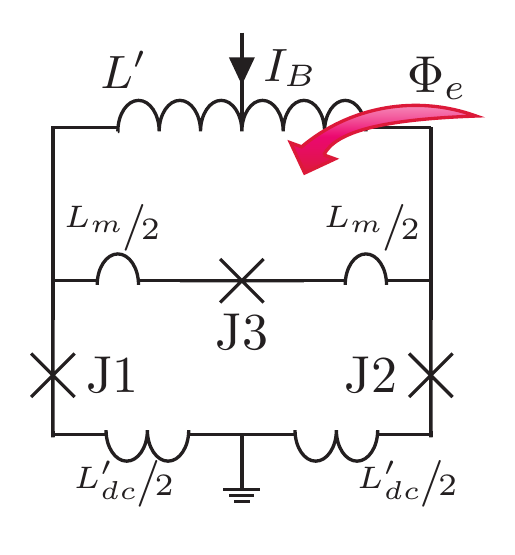}}\hfill	
\subcaptionbox{}{\includegraphics[width=8cm]{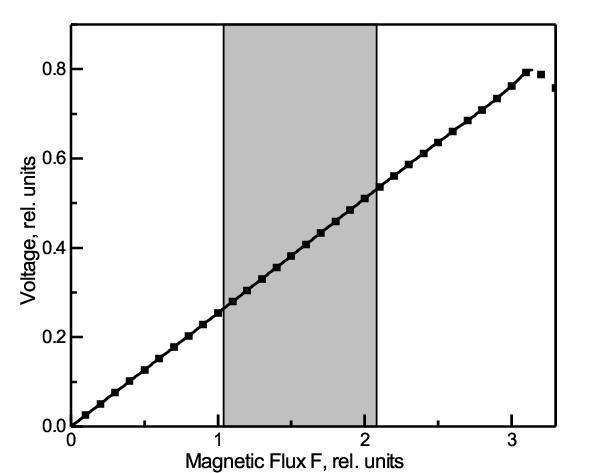}}\\
\subcaptionbox{}{\includegraphics[width=8cm]{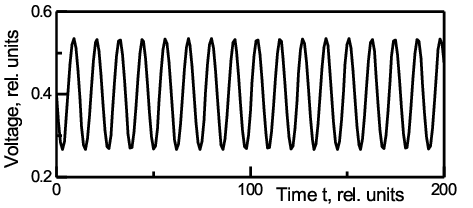}}\hfill	
\subcaptionbox{}{\includegraphics[width=8cm]{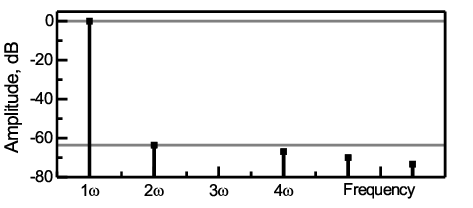}}
\caption{\label{fig:Example}\textit{(a)} Schematic of bi-SQUID. J1, J2, J3 is Josephson junctions, $L=L'+L_m$ and $L_{dc}=L_{dc}'+L_m$ are inductance of SQUIDs' loops, the input signal, flux $\Phi_{ex}$, is applied to the only rf SQUID (top) loop. From \cite{Kornev2020_SUST} with permission. \textit{(b-d)} Workflow of method for analyzing linearity of bi-SQUID: \textit{(b)} simulated voltage-flux characteristic $V_\text{signal}(F)$ (points, only every thousandth point is shown) and an interpolated one $\mathfrak{V}_\text{signal}(F)$ (line), the gray area indicates the used part of the signal; \textit{(c)} resulted signal $\mathfrak{V}_{one-tone}(t)$ in time-domain; \textit{(d)} amplitudes of harmonics as a result of FFT analysis, amplitudes are given in decibels and are normalized so that the amplitude of the fundamental harmonic is $0\text{ dB}$. Resulting linearity is $63\text{ dB}$.}
\end{figure}

On the figure \ref{fig:Example} workflow of method for analyzing linearity of example device (bi-SQUID \cite{Kornev2020_SUST}) are shown. Bi-SQUID is modified dc SQUID, which contains an additional, third Josephson junction, connected in parallel with the main loop inductance of the reference dc SQUID as shown in figure \ref{fig:biSQUID}.

Resulting linearity of the example device is $63\text{ dB}$.

\end{document}